\documentclass[prl,twocolumn]{revtex4}
\usepackage{bm}
\usepackage{graphicx}
\usepackage{amssymb}
\usepackage{amsmath}
\usepackage{eufrak}
\usepackage{color}
\usepackage[utf8]{inputenc} 
\usepackage{hyperref}
\usepackage{pifont}
\usepackage{ulem}

\newcommand{\nix}[1]{}

\begin{document}

\title{Room temperature high frequency transport of Dirac fermions\\
in epitaxially grown Sb$_2$Te$_3$ based topological insulators}
\author{P.~Olbrich$^1$, L.\,E.~Golub$^2$, T.~Herrmann$^1$, S.\,N.~Danilov$^1$,
H.~Plank$^1$,  V.\,V.~Bel'kov$^2$, G.~Mussler$^3$,
Ch.~Weyrich$^3$, C.\,M.~Schneider$^3$,   J.~Kampmeier$^3$,
D.~Gr\"{u}tzmacher$^3$, L.~Plucinski$^3$, M.~Eschbach$^3$, and
S.\,D.~Ganichev$^1$ }
\affiliation{$^1$  Terahertz Center, University of Regensburg, 93040 Regensburg, Germany}
\affiliation{$^2$Ioffe Physical-Technical Institute,
194021 St.~Petersburg, Russia}
\affiliation{$^3$
J\"ulich Aachen Research Alliance (JARA-FIT),
Leo Brandt Stra{\ss}e, 52425 J\"ulich, Germany}

\begin{abstract}
We report on the observation of photogalvanic effects in
epitaxially grown Sb$_2$Te$_3$ three-dimensional (3D) topological
insulators (TI). We show that asymmetric scattering of Dirac
electrons driven back and forth by the terahertz electric field
results in a \textit{dc} electric current. Due to the ``symmetry
filtration'' the \textit{dc} current is generated in the surface
electrons only and provides an opto-electronic access to
 probe the electric transport in TI, surface
domains orientation and details of electron scattering  even in
3D TI at room temperature where conventional surface electron
transport is usually hindered by the high carrier density in the
bulk.
\end{abstract}


\maketitle{}

A new state of matter called the topological insulator has recently
been theoretically predicted and experimentally observed in a number of
materials, such as Bi$_2$Se$_3$, Sb$_2$Te$_3$,
and Bi$_2$Te$_3$, for reviews
see~\cite{introd_1,introd_2,introd_3}.
The main feature of TI emerges from  its band structure.
While the bulk of TI is an insulator with an inverted band structure
its surface hosts gapless states with a linear energy dispersion.
Thus, carriers at the surface of TI are expected to have
unique properties,  such as extremely high
mobilities or  a spin-locked transport behavior,
and TI are considered to be prospective for novel applications in
the field of spintronics, optoelectronics, or quantum computing.
Hence, a fabrication of TI materials and, in particular,
study of their transport properties moved into the focus
of current research. However, in almost all known 3D TI the
\textit{dc} electron transport is often hindered by the high
carrier density in the bulk~\cite{bulk_states1,bulk_states2,bulk_states3}.
A promising way to overcome this problem serves
the recent progress in growth of 3D TI applying molecular-beam-epitaxy (MBE)
technique, see e.g.,~\cite{Borisova2012,Plucinski_2013}.
The existence of TI surface states in such materials has been
demonstrated by the angle-resolved-photoemission-spectroscopy (ARPES)~\cite{ARPES_1,ARPES_2,ARPES_3}.
Furthermore, low-temperature electric measurements
in thin films and nanowires indicate substantial 
surface state transport~\cite{transp1,transp2,transp3,transp4,transp5,transp6,newreview}.
However, the electron transport exclusively determined by surface electrons,
in particular, at room temperature, remains a challenge.

Here we report on the observation and study of a room temperature
high frequency transport phenomena solely determined by 2D Dirac
fermions in 3D TI. We show that  excitation of MBE-grown
Sb$_2$Te$_3$ crystals by terahertz (THz) electric fields results
in a photogalvanic effects (PGE): a nonlinear transport effect
yielding a  \textit{dc} electric current proportional to the
square of the \textit{ac} electric
field~\cite{book_Ivchenko,book}. A selective excitation of
\textit{dc} current in TI surface states becomes possible due to
the specific feature of PGE, whose prerequisite is a lack of
inversion center. As Sb$_2$Te$_3$ crystals, like  most of the 3D
TI, is centrosymmetric, this requirement is fulfilled for the
surface states only. Due to this ``symmetry filtration'', the PGE
is generated in the surface electron system only, even in the
materials with substantial conductance in the bulk. We demonstrate
that the PGE is caused by asymmetric scattering of Dirac electrons
driven back and forth by the
THz 
field. The effect reflects the
surface symmetry and allows one to determine the orientation
of the surface domains, to probe high frequency conductivity in TI,
and to study tiny details of electron scattering.

\begin{figure}[t]
\includegraphics[width=0.99\linewidth]{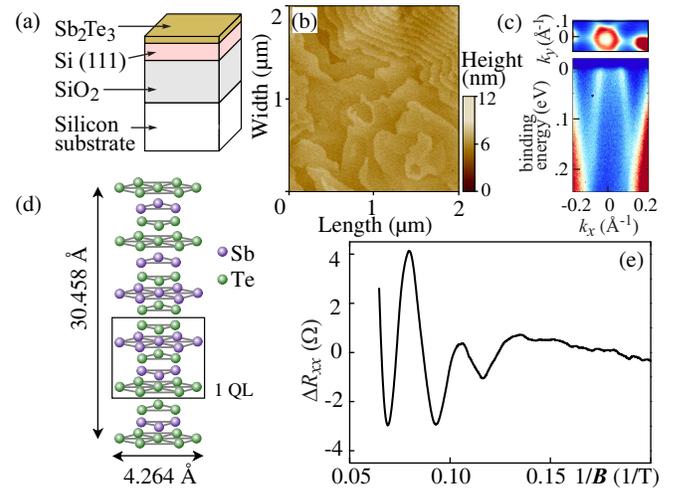}
\caption{
(a) Sample sketch,
(b) AFM surface scan
(c) ARPES measurement showing  \textit{p}-type linear dispersion,
(d) Structure of Sb$_2$Te$_3$ layer,  and
(e) SdH oscillations.
}
\label{figure01}
\end{figure}

\begin{figure}[t]
\includegraphics[width=0.9\linewidth]{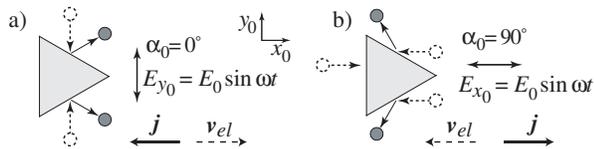}
\caption{
Model of the PGE excited in Dirac fermions of Sb$_2$Te$_3$ TI
due to asymmetry of elastic scattering by  wedges.
}
\label{wedges}
\end{figure}

Before discussing the experimental results we address the basic
physics of the PGE in 3D TI and set requirements to
the experimental geometry. The surface states of TI are based on
the crystalline structure, see Fig.~\ref{figure01}~(d),
which includes a sequence of five atomic layers, so
called a quintuple layer (QL),  oriented perpendicularly to
the \textit{c}-axis~\cite{NatPhys09}.
The point group symmetry of Sb$_2$Te$_3$ bulk crystal is D$_{\rm 3d}$,
which includes the center of inversion, whereas the
\textit{surface} lacks
the space inversion and its point group is C$_{\rm 3v}$.
The trigonal symmetry of 2D surface carriers
makes the elastic scattering asymmetric
giving rise to a \textit{dc} electric current
in response to \textit{ac} electric field.
The process of current generation is illustrated
in Fig.~\ref{wedges}
where the scatterers are sketched as randomly distributed but identically
oriented wedges lying in the QL-plane. In the absence of radiation, the flows
of anisotropically scattered
electrons exactly compensate each other.
Application of the
linearly polarized THz field results in \textit{alignment}
of electron momenta: the total number of Dirac electrons
driven back and forth by \textit{ac} electric field $\bm{E}(t)$ increases
while the number of particles moving, e.g.  perpendicularly  to the field direction, decreases.
The corresponding stationary correction
to the electron distribution function scales as a
square of the $ac$ electric
field magnitude~\cite{alignment}.
The alignment of electron momenta itself does not lead to the
\textit{dc}  electric current but, due to asymmetric scattering by wedges,
the excess of the number of carriers moving
along the field violates
the balance of the flows~\cite{Belinicher-Strurman-UFN,Sturman-Fridkin-book,GaN_LPGE}, and
the linear PGE current is generated. The resulting  current direction depends on the relative
orientation of the \textit{ac} electric field and wedges:  e.g. the field parallel to the wedges base,
see Fig.~\ref{wedges}~(a),
yields the current flowing in $x_0$-direction
while rotation of the electric field by  90$^\circ$
reverses the direction of the current, see Fig.~\ref{wedges}~(b).
Symmetry analysis yields the
polarization dependence of the PGE current density $\bm j$:
\begin{align}
\label{phenom}
    j_{x_0} =   \chi(|E_{x_0}|^2-|E_{y_0}|^2) = - \chi|E_0|^2 \cos 2 \alpha_0, \\
    j_{y_0}= -\chi(E_{x_0} E_{y_0}^*+E_{y_0}E_{x_0}^*) = \chi |E_0|^2 \sin 2 \alpha_0. \nonumber
\end{align}
Here  $E_0$ is the electric field amplitude, the factor $\chi$ is the
single linearly-independent constant, and $\alpha_0$ is counted anti-clockwise from $y_0$.
Note that the brackets in the first and the second equations divided by $|E_0|^2$ represent 
the Stokes parameters~\cite{Born_Wolf}  of the linearly-polarized radiation 
$s_1$ and $s_2$, respectively.

\begin{figure}[t]
\includegraphics[width=0.99\linewidth]{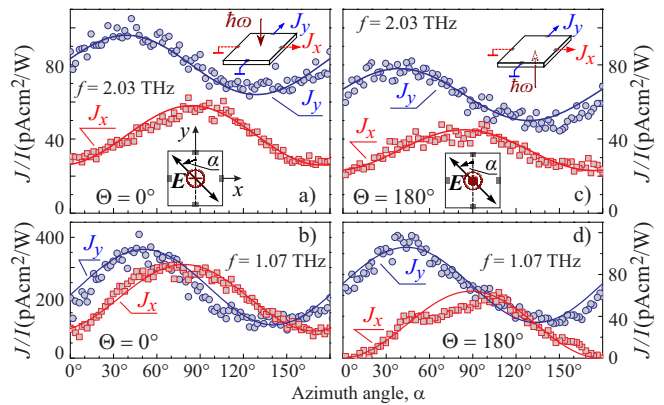}
\caption{ Normalized photocurrents $J_x/I$ and $J_y/I$
for front (left panels) and back (right panels)
illumination of the Sb$_2$Te$_3$ sample at room temperature.
Solid lines show fits  after Eqs.~\eqref{fit_alpha2}.
Insets sketch the setups.
}
\label{figure02}
\end{figure}

In the bulk centrosymmetric Sb$_2$Te$_3$ crystals spatial inversion forbids the
linear coupling between the current and electric field square
and, in contrast to the surface, the PGE \textit{dc} current cannot be generated.
This difference has been addressed in Ref.~\cite{mciver}, where control over
TI photocurrents with light polarization was demonstrated
by study of photon helicity-induced (circular) PGE. However,
the strict ``symmetry filtration'' is violated due to a possible competing
contribution of the photon drag effect~\cite{book_Ivchenko,book} which does not
require the lack of inversion symmetry. While in analysis of
Ref.~\cite{mciver}
the photon drag effect in the bulk  has been ruled out due to its spin-degeneracy
most recent observations demonstrated that substantial linear and circular
photon drag currents can be efficiently generated even in 2D material with vanishing spin-orbit
coupling, such as graphene,  and gives a response comparable with
PGE~\cite{book_Ivchenko,book,GlazovGanichev_review}. A straightforward way to distinguish
the PGE response emerging from the surface states and photon drag effect
provides experiments with reversed direction of the light propagation. Indeed while the
PGE is determined by the electric field orientation and is insensitive to the
radiation propagation direction
the photon drag current being proportional to
the photon momentum ${\bm q}$
\begin{equation}
\label{PDE}
    j_{x_0} =   {\cal T} |E_0|^2 q_z s_1,  \qquad j_{y_0}=  - {\cal T} |E_0|^2 q_z s_2,
\end{equation}
reverses its sign. Here $ {\cal T}$ is the photon drag constant and $z \parallel [111]$.
Note that both types of the photocurrent behave identically upon variation
of the radiation polarization state, cf. Eqs.~\eqref{phenom}
and~\eqref{PDE}. This indistinguishable behavior
is also obtained for linear and circular photocurrents
excited at oblique incidence, see Supplementary Material.
The only way to separate the surface and the bulk transport is to
excite photocurrents applying the radiation from both sides of the sample.

To explore the high frequency transport in Dirac fermion systems
we studied photocurrents excited by THz radiation in
Sb$_2$Te$_3$ grown by MBE on Si(111) wafers.
A corresponding sketch of the structure is shown in
Fig.~\ref{figure01}~(a) and~(d) and details  of sample
growth~\cite{Plucinski_2013} are
given in Supplementary Materials.
At low temperatures ($T \approx 30$~K) the samples have
a mobility of~290\,cm$^{2}$/Vs
and
$p$-type carrier density about\,5\,$\times$\,10$^{19}$\,cm$^{-3}$.
Values are obtained by standard transport measurements, see
Fig.~\ref{figure01}~(e) showing clear pronounced Shubnikov-de~Haas
(SdH) oscillations. The band gap $E_g$ is in the order of 170~meV
with the Fermi level referred to the Dirac point $\varepsilon_{\rm
F} \approx 65$~meV and a corresponding Fermi velocity, $v_0 = 4.36
\times 10^5$~ m/s, measured by scanning tunneling spectroscopy and
ARPES, see Fig.~\ref{figure01} (c).
These results show that the sample should emerge
topological properties, e.g. conducting surface states~\cite{Plucinski_2013}.
To measure current a pair of ohmic contacts ($R \approx 400~\Omega$)
was centered on opposite edges of the squared shaped sample.
To apply an $ac$ electric field ${\bm E}(t)$ in the plane of QL we used
a normally incident linearly polarized THz radiation of molecular
laser~\cite{Karch2010,Ganichev02,Ziemann2000}, see insets in
Fig.~\ref{figure02} and Supplementary Materials
for details. The \textit{ac} field direction was rotated by
an azimuth angle $\alpha$ in respect to a sample edge defined
as $y$-axis. The angle of incidence $\Theta$
for front and back illuminations were $0^\circ$ and $180^\circ$, respectively.

Exciting Sb$_2$Te$_3$ samples with the THz electric field
we observed a {\it dc} current exhibiting
a characteristic polarization dependence shown in Fig.~\ref{figure02}.
Panels (a) and (b) present the photocurrents, $J_x(\alpha)$ and $J_y(\alpha)$, excited
by front illumination and measured as a function of the $ac$ electric field orientation.
The signals are well fitted by
\begin{align}
\label{fit_alpha2}
    J_x = [- A(f) \cos (2\alpha - 3\Phi) +C(f)] I, \\
    J_y = [ A(f) \sin (2\alpha- 3\Phi) + C'(f)] I,
\end{align}
where $A$, $C$, $C'$ and $\Phi$ are fitting parameters,
and $I \propto E_0^2 $ is the radiation intensity.
The photocurrent for the back excitation ($\Theta = 180^\circ$)
is shown in Fig.~\ref{figure02}~(c)~and~(d).
As an important result we obtained that
the sign of the current as well as its dependence on the
azimuth angle $\alpha$ remains unchanged.
The same result is obtained for several other samples
grown in the similar way (not shown). While in all samples
the polarization dependence  for front and back
illumination remains unchanged
the phase shift, being constant for each sample,
varies from -4$^\circ$ to -10$^\circ$.

\begin{figure}[t]
\includegraphics[width=\linewidth]{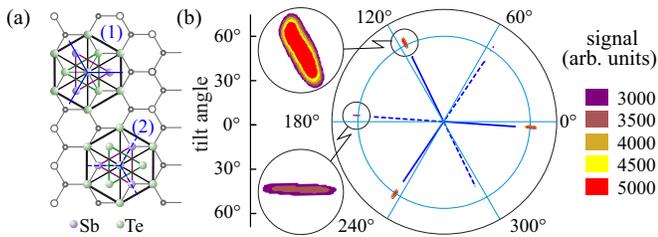}
\caption{ (a) Two possible orientations of the domains. (b) Result of $X$-ray
diffraction measurements on the Sb$_2$Te$_3$ sample, showing that one domain orientation dominates.}
\label{figure04}
\end{figure}

\begin{figure}[t]
\includegraphics[width=0.99\linewidth]{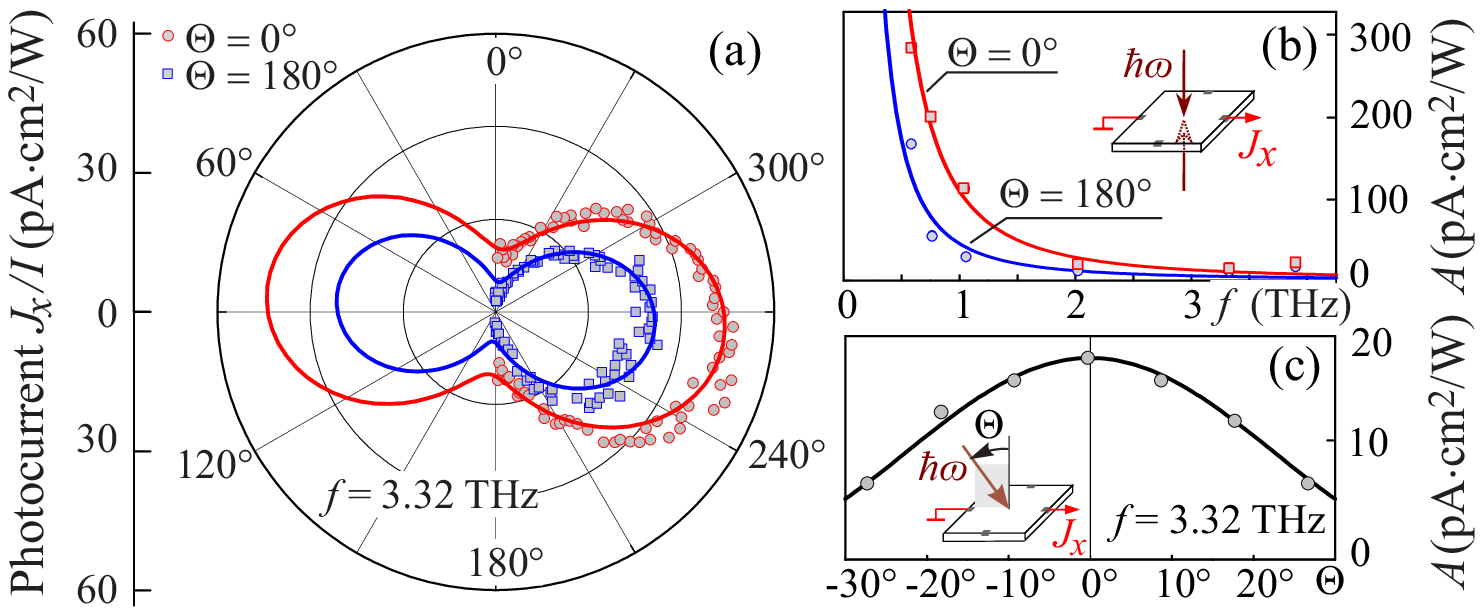}
\caption{(a) Photocurrents $J_x(\alpha)/I$
measured for front and back illumination at $T=296$\,K.
Lines are fits after Eq.~\eqref{fit_alpha2} with $\Phi = -4^\circ$.
Coefficient $A$
as a function of  frequency $f$ (b) and  the angle
$\Theta$ (c). Lines in (b) show fits after $A \propto 1/f^2$.
}
\label{figure03}
\end{figure}

Figure~\ref{figure02} shows that besides the offsets $C$ and $C'$~\cite{footnote},
the functional behavior of the photocurrents follows 
Eqs.~\eqref{phenom} and~\eqref{PDE}
extended by a phase shift given by the angle
$3\Phi$ and using  $\alpha = \alpha_0 + \Phi$, see Supplementary Material. The fact that the
sign of the coefficient $A$ remains unchanged at ${\bm q} \rightarrow - {\bm q}$ inversion
unambiguously demonstrate that the photocurrent is dominated by the
photogalvanic effect and, consequently,
is generated in 2D Dirac fermions.
Symmetry analysis of the photogalvanic effect shows that the $3\Phi$ phase shift
comes from the misorientation of $y_0$-direction
and the sample edge denoted as $y$ axis,
which in Eq.~\eqref{phenom} is assumed to be zero.
The phase shift $3\Phi$ obtained for the arbitrary 
orientation wedges takes into account both the orientation of the $ac$
electric field with respect to $y_0$ and the fact that the
currents $j_x$ and $j_y$ are probed at angle $\Phi$  with respect 
to $x_0$ and $y_0$, respectively.
While being differently aligned with respect to the sample edges,
all measurements reveal an almost uniform orientation of scattering
centers characterized by the three-fold symmetry.
The same result is obtained applying $X$-ray diffraction measurements.
While two possible types of domains
can be formed during the growth of Sb$_2$Te$_3$ on (111)-oriented Si substrate,
see Fig.~\ref{figure04}~(a), the $X$-ray data shown in Fig.~\ref{figure04}~(b)
demonstrate that
the majority of the domains has the same orientation.
As an important result, the angle $\Phi = -4^\circ$ obtained
from the photocurrent measurements is equal to
that measured by $X$-ray diffraction. This is clearly seen from
comparison of Fig.~\ref{figure04}~(b) and Fig.~\ref{figure03}~(a),
which shows the photocurrent $J_x(\alpha)/I = [-A(f) \cos (2\alpha + 12^\circ) +C]$
for front and back illumination in polar coordinates.

Varying the electric field frequency we obtained that the
parameter $A(f)$, which determines the photocurrent magnitude,
strongly increases with the frequency decrease.
Figure~\ref{figure03}~(c) shows that at low frequencies
it closely follows the law $A \propto 1/f^2$.
This figure also reveals that PGE for back irradiation
is weaker than that for the front one.
We attribute this result to either absorption in the substrate and/or
to the contribution of PGE excited in the surface states at the  
SbTe/substrate interface, which can yield a different
strength of the photoresponse of Dirac fermions.

Applying radiation at oblique incidence we found that the
photocurrent given by the coefficients $A(f)$ slightly
decreases with  increasing the angle of incidence
$|\Theta|$, see Fig.~\ref{figure03} (b).
This fact indicates that in the described experimental arrangement the
dominant contribution to the current stems from the PGE at normal incidence
and other possible contributions of PGE or photon drag effect which, 
as addressed above, can be exited at oblique incidence
do not play any substantial role. Additional experiments using
circularly polarized light also revealed that the circular
photocurrent is either absent  or  hidden by the polarization
independent photocurrent.

While the explanation of the photogalvanic effect
has been given in a pictorial way above, we resort
now to a microscopic description based on the Boltzmann
kinetic equation for the electron distribution function $f_{\bm p}(t)$
\begin{equation}
    {\partial f_{\bm p} \over \partial t} + e \bm E \cdot {\partial f_{\bm p} \over \partial \bm p}
    = - \sum_{\bm p'} \left( W_{\bm p', \bm p} f_{\bm p} - W_{\bm p, \bm p'} f_{\bm p'} \right),
\end{equation}
where  the electric field ${ \bm E(t) = E_0 \exp{(-{\rm i}\omega t)} + c.c. }$,
and $W_{\bm p', \bm p}$ is a probability for an electron to have the momenta
$\bm p$ and $\bm p'$ before and after scattering, respectively.
Lack of inversion center for the surface electrons makes their elastic scattering
asymmetric:  even for  isotropic
scatterers,  $W_{\bm p, \bm p'} \neq W_{-\bm p, -\bm p'}$~\cite{Belinicher-Strurman-UFN,Sturman-Fridkin-book}
and is given by
\begin{equation}
    W_{\bm p', \bm p} = W^{(s)}_{\bm p', \bm p} + W^{(a)}_{\bm p', \bm p},
\end{equation}
where $W^{(s)}_{\bm p', \bm p}=W^{(s)}_{-\bm p', -\bm p}$ is the symmetric part, and
the scattering asymmetry is described by ${W^{(a)}_{\bm p', \bm p}=-W^{(a)}_{-\bm p', -\bm p} }$.
The absence of backscattering for Dirac fermions is taken into consideration by the standard factor in the symmetrical part~\cite{backscattering}:
$W^{(s)}_{\bm p', \bm p} \propto \cos^2[({\varphi_{\bm p'} - \varphi_{\bm p}) /2 ]}$,
where $\varphi_{\bm p},\varphi_{\bm p'}$ are the polar
angles of the 2D vectors $\bm p$ and $\bm p'$.

We iterate the kinetic equation in the second order in the field amplitude
taking into account that
$eE_0v_0\tau_{\rm tr}/\varepsilon_{\rm F} \ll 1$,
where the linear 2D energy dispersion with the velocity
$v_0$ is considered, and the transport scattering time, $\tau_{\rm tr}$, determining the mobility
of 2D Dirac fermions is
related to the symmetric part of the scattering probability as
${\tau_{\rm tr}^{-1}=\sum_{\bm p'}W^{(s)}_{\bm p', \bm p}[1-\cos{(\varphi_{\bm p'} - \varphi_{\bm p}})]}$.
As a result, we find the stationary correction to the distribution function
$\delta f_{\bm p} \propto |E_0|^2$~\cite{alignment}.
Calculating the photocurrent density by the standard expression  ${\bm j = e\sum_{\bm p}\delta f_{\bm p} v_0\bm p/p}$,
we obtain
\begin{align}
    \label{chi}
j_{{x_0},{y_0}} = & \pm  s_{1,2} |E_0|^2\,
e v_0 \sigma(\omega) \\
&\times \left[
    {1\over \varepsilon_{\rm F}^2} {d ({\Xi}\tau_2\varepsilon_{\rm F}^2) \over d \varepsilon_{\rm F}}
    +{1-\omega^2\tau_{\rm tr}\tau_2\over 1+(\omega\tau_2)^2}
     {{\Xi}\tau_2 \varepsilon_{\rm F}\over \tau_{\rm tr}} {d(\tau_{\rm tr}/ \varepsilon_{\rm F})\over d \varepsilon_{\rm F}}
    \right].\nonumber
\end{align}
Here the high-frequency conductivity is given by the Drude expression
${ \sigma(\omega)= n_s(ev_0)^2\tau_{\rm tr}/[\varepsilon_{\rm F} (1+\omega^2\tau_{\rm tr}^2)] }$
with $n_s$ being the concentration of 2D carriers which are degenerate, and $+s_1$, $-s_2$ correspond to $j_{x_0}$, $j_{y_0}$, respectively.
The time $\tau_2 $ being of the order of $\tau_{\rm tr}$ describes relaxation of the above discussed alignment of electron momenta.
It is given by ${\tau_2^{-1}=\sum_{\bm p'}W^{(s)}_{\bm p', \bm p}[1-\cos{2(\varphi_{\bm p'} - \varphi_{\bm p}})]}$.
The scattering asymmetry is taken into account by the factor ${\Xi \ll 1}$
\begin{equation}
    {\Xi} = \tau_{\rm tr} \sum_{\bm p'}
    \left< 2\cos{\varphi_{\bm p}} \cos{2\varphi_{\bm p'}} \, W^{(a)}_{\bm p', \bm p} \right>_{\varphi_{\bm p}},
\end{equation}
where the brackets denote averaging over the directions of $\bm p$ at the Fermi circle.
Note that Eq.~\eqref{chi} agrees with Eqs.~\eqref{phenom} obtained
from the phenomenological arguments.

The microscopic theory of the PGE presented above
describes all major features observed in the experiments.
It shows that the $dc$ electric currents probed along and
normal to the wedges base are proportional to the square of the
$ac$ electric field amplitude $|E_0|^2$, described by one constant  $ A(f) I = \chi |E_0|^2$
and vary upon rotation of the electric field direction after the Stokes
parameters $s_1$ and $s_2$, respectively.
Such a behavior is observed in the experiment, see Fig.~\ref{figure02}.
Then, the experimental data show that the photocurrent scales as
$1/\omega^2$ at low frequencies, see Fig.~\ref{figure03}~(b). 
This behavior follows from the theory: at $\omega\tau_{\rm tr},\omega\tau_2 \gg 1$
from Eq.~\eqref{chi} we obtain $j \sim e^3v_0\Xi |E_0|^2/(\hbar^2\omega^2)$.
A small deviation from this behavior observed at high frequencies,  see Fig.~\ref{figure03}~(b),
can be caused by the surface roughness, see  Fig.~\ref{figure01}~(b),
which modifies the frequency behavior of $\sigma(\omega)$, like it has been reported
for the epitaxial graphene~\cite{Kuzmenko}
and other multilayer thin film systems, see e.g.,~\cite{Drude}.
Equation~\eqref{chi} also reveals that the
magnitude of the PGE current, as well as its
functional behavior upon variation of the electric field
frequency or temperature,
are determined by the dominant elastic scattering
mechanism for Dirac fermions. In particular, the photocurrent
can be generated for the scattering by Coulomb impurities or phonons but
vanishes for that by short-range impurities, see Supplementary Materials.
Thus, the observation of the PGE in Sb$_2$Te$_3$ crystals
indicates the dominant role of the Coulomb scattering in the
surface state electron transport.
Furthermore, tiny details of scattering can be obtained from the
study of the PGE spectral behavior at $\omega\tau_{\rm tr} \leq
1$, where the scattering mechanism affects the frequency
dependence of the photocurrent (via $\tau_{\rm tr}$ and $\tau_2$).
Finally, we estimate the PGE magnitude following Eq.~\eqref{chi}.
We obtain for the radiation of $f = 0.6$~THz focused in 2.8~mm spot
the experimental  value \mbox{$A = 280 $~pA cm$^2$/W}
for the scattering asymmetry factor ${\Xi} \sim 10^{-5}$.
The latter value is consistent with the  theoretical estimations, see
Supplementary Materials.

To summarize, the observed photogalvanic effect in
the surface states  provides an opto-electronic
method to selectively excite and study high frequency
transport of the Dirac fermions in 3D TI. The photocurrent, being sensitive 
to the surface symmetry and scattering details, 
can be applied to map the domain orientation in 3D TI
 and study the high frequency
conductivity of the surface states even at room temperature.
Finally we note that while in the studied frequency range
and materials the photoresponse is dominated by
PGE our analysis demonstrates, that to ensure
that the photoresponse comes from the Dirac fermions
and to exclude a possible contribution of the
bulk, the experiments with front and back sample excitation are required.

\acknowledgments We thank M.\,M. Glazov, M.\,Schmalzbauer,  E.\,L.\,Ivchenko and 
J.\,H.\, Bardarson
for fruitful discussions. The  support from the DFG (SPP 1666), HGF virtual Institute,
RFBR and EU programme POLAPHEN is gratefully acknowledged.


\begin{thebibliography}{99}
\bibitem{footnote} The polarization independent offset may be caused by
the non perfectly flat surface, see the AFM photograph in Fig.~\ref{figure01},
which locally reduces the symmetry of the surface states and allows a
polarization independent
contribution to the photogalvanic effect.
\end{thebibliography}
\end{document}


\title{Supplementary material for ``Room temperature high frequency transport of Dirac fermions
in epitaxially grown Sb$_2$Te$_3$ based topological insulators''}
\author{P.~Olbrich$^1$, L.\,E.~Golub$^2$, T.~Herrmann$^1$, S.\,N.~Danilov$^1$,
H.~Plank$^1$,  V.\,V.~Bel'kov$^2$, G.~Mussler$^3$,
Ch.~Weyrich$^3$, C.\,M.~Schneider$^3$,   J.~Kampmeier$^3$,
D.~Gr\"{u}tzmacher$^3$, L.~Plucinski$^3$, M.~Eschbach$^3$, and
S.\,D.~Ganichev$^1$ }
\affiliation{$^1$  Terahertz Center, University of Regensburg, 93040 Regensburg, Germany}
\affiliation{$^2$Ioffe Physical-Technical Institute,
194021 St.~Petersburg, Russia}
\affiliation{$^3$
J\"ulich Aachen Research Alliance (JARA-FIT),
Leo Brandt Stra{\ss}e, 52425 J\"ulich, Germany}

\maketitle

\section{Sample preparation}

Our Sb$_2$Te$_3$ samples were grown by molecular beam epitaxy (MBE)
on Si(111) wafers.
Prior to the deposition, the Si substrates were chemically
cleaned by the HF-last RCA procedure to remove
the native oxide and passivate the surface with hydrogen.
The substrates were subsequently heated
in-situ to 600$^\circ$C for 20 min to desorb the hydrogen
atoms from the surface. The Sb and Te
material fluxes were generated by effusion cells that were
operated at temperatures of 450$^\circ$C (Sb)
and 380$^\circ$C (Te). The Te shutter was opened 2 seconds
before the Sb shutter, in order to saturate
the Si substrate surface with Te.
Throughout the growth, the substrate temperature was set at 300$^\circ$C.
A thickness of 27~nm
Sb$_2$Te$_3$ were deposited at a slow growth rate of 9 nm/h in order
to suppress domain formation
and to obtain a smooth and uniform sample surface.

\section{Photocurrent measurements}

To excite photocurrent we used alternating
electric field $\bm{E}(t)$ of terahertz radiation of
TEA-CO$_2$ laser pumped terahertz molecular laser)~\cite{book,Schneider2004,Karch2010}.
The operating parameters of laser lines used in experiment are given in Table~\ref{rgblines}.
The applied single pulses had a duration of about 100~ns and repetition rate of 1~Hz.
The radiation power was controlled by the THz photon drag detector~\cite{Ganichev84p20}.
The THz field was focused on a typical spot diameter of about 1 to 3~mm,
being smaller than the sample size of $5 \times 5$~mm$^2$.
The spatial beam distribution had an almost Gaussian profile which was
measured by a pyroelectric camera~\cite{Ziemann2000p3843}.
The electric field amplitude of incoming radiation
was varied from about $1$ to $30$~kV/cm (radiation intensities from
$\approx 1$ to 1000~kW/cm$^2$).
The \textit{dc} current in response to terahertz electric field  was measured
in the unbiased samples by the voltage drop across a 50\,$\Omega$ resistor.
The signal was recorded with a storage oscilloscope.
The initial  laser radiation was linearly polarized along the $y$-axis.
To vary the radiation polarization, $\lambda$/2 and $\lambda$/4
crystal quartz plates were employed.
By applying the $\lambda/2$ plates, we varied the azimuth angle $\alpha$
between the polarization plane of the radiation incident upon the sample
and the $y$ axis. By applying $\lambda$/4 plates, we obtained
elliptically (and circularly) polarized radiation. In this case, the polarization state
is determined by the angle $\varphi$
between the plate optical axis and the incoming laser polarization with electric field vector along
the $y$ axis. In particular, the radiation helicity is given by
$P_{\rm circ} = \sin{2 \varphi}$~\cite{Ganichev02}.

%
\begin{table}[t]
\caption{Characteristics of  terahertz laser   lines used in experiments.}
\begin{center}
\begin{tabular}{cccccc}
\hline\\[-6pt]
&
&
& Line of CO$_2$ & Max. intensity &
\\[3pt]
$f$ (THz) & $\lambda$~($\mu$m) & $\hbar \omega$ (meV) & pump laser & (kW cm$^{-2}$) & Medium \\[3pt]
\hline \\[-6pt]
  8.57  &  35    &  35.3  & 10P(24)   &  300             & NH$_3$ \\[3pt]
  3.95  &  76    &  16.3  & 10P(26)   &   10             & NH$_3$ \\[3pt]
  3.32  &  90.5  &  13.7  &  9R(16)   & 1000             & NH$_3$ \\[3pt]
  2.03  & 148    &  8.4   &  9P(36)   &  790             & NH$_3$ \\[3pt]
  1.07  & 280    &  4.4   & 10R(8)    &   50             & NH$_3$ \\[3pt]
  0.78  & 385    &  3.2   &  9R(22)   &   10             & D$_2$O \\[3pt]
  0.61  & 496    &  2.5   &  9R(20)   &    8             & CH$_3$F \\[3pt]
\hline
\end{tabular}
\end{center}
\label{rgblines}
\end{table}

\section{Photogalvanic current in the rotated coordinate frame}

\begin{figure}[t]
\includegraphics[width=0.6\linewidth]{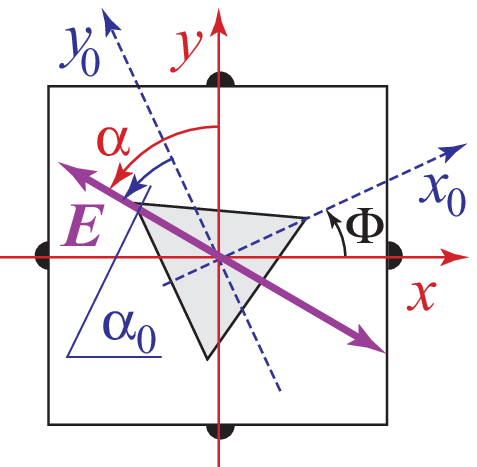}
\caption{ Schematic representation of the orientation of the 
crystallografic axes $x_0$, $y_0$ tilted by the angle $\Phi$
relative to the directions $x$, $y$ along which the current is measured.
Double arrow sketches the $ac$ electric field vector $\bm E$.
Azimuth angles $\alpha$ and $\alpha_0$ are counted from the axes 
$y$ and $y_0$, respectively. 
}
\label{fig:Phi-angle}
\end{figure}

The photogalvanic current direction is determined by the orientation of 
the electric vector in respect to crystallographic axes. 
Therefore, most convenient for this effect would be to investigate the current 
in the directions along and perpendicular to one of the mirror reflection planes 
of the C$_{3 {\rm v}}$ point group.
However, 
in real samples the direction of contacts may not coincide with the crystallographic axes.
Therefore, we present here the equations describing photogalvanic current 
and its polarization dependence for the arbitrary orientation of contacts 
in respect to the crystallographic directions.  
To analyze this we introduce the angle $\Phi$ between 
the in-plane directions $x$ and $y$ along which the current is probed 
and the crystallographic axes $x_0$ and $y_0$, see Fig.~\ref{fig:Phi-angle}.
The symmetry consideration yields the expression for 
$x$ and $y$ components of the photogalvanic current excited 
at normal incidence:
\begin{eqnarray}
    \label{j_normal'}
&&j_{x} =  -\chi |E_0^2|  \: \cos{(2 \alpha - 3\Phi)} \, , \\
&&j_{y} =  \chi |E_0^2|  \: \sin{(2 \alpha - 3\Phi)}  \, , \nonumber
\end{eqnarray}
where $\alpha$ is counted from $y$, i.e. ${\bm E}(\alpha=0) \parallel y$.
For $\Phi = 0$ the azimuth angle $\alpha = \alpha_0$ and the above equation reduces to 
Eqs.~(1) of the main manuscript.
The phase shift $3\Phi$ obtained for the arbitrary 
orientation wedges takes into account both the orientation of the $ac$
electric field with respect to $y_0$ and the fact that the
currents $j_x$ and $j_y$ are probed at angle $\Phi$  with respect 
to $x_0$ and $y_0$, respectively.

\section{Phenomenological theory of photocurrents in TI}

While the paper is devoted to photogalvanic effect excited at normal incidence 
here, for completeness, we give the expressions for the photogalvanic and photon drag 
currents excited by oblique-incident radiation.
We treat these nonlinear effects without going into
microscopic details making use the symmetry arguments. This approach
allows one to conclude on the experimental geometry and conditions of observation of the
effect under consideration as well as  to describe its variation with change
of macroscopic parameters, such as intensity of the radiation, 
its polarization and angle of incidence without knowing of 
the microscopic origin. We consider the  photocurrents generated in the 
surface states (point symmetry  C$_{\rm 3v}$)
and in the centrosymmetric bulk material (point group D$_{\rm 3d}$).

In the following we focus on the photocurrent contributions which can be generated 
at oblique incidence only. Note that the normal-incident radiation induced
contributions of the PGE and photon drag effect described in the main text
remain at oblique incidence.
In general two kinds of the photocurrent can be generated at oblique-incident radiation
\begin{equation}
	\bm j^{(oblique)} = \bm j^{(surf)} + \bm j^{(drag)},
\end{equation}
where $\bm j^{(surf)}$ caused by PGE in TI surface  and  $\bm j^{(drag)}$ is due to the photon drag effect where contributions from both surface and bulk are present which are phenomenologically identical in our system.

The photocurrent  $\bm j^{(surf)}$ is a typical contribution for any inversion-asymmetric two-dimensional system:
\begin{align}
	&	j_x^{(surf)} =  \chi_1 (E_xE_z^*+E_zE_x^*) + \gamma P_{\rm circ} |E_0|^2 {q_y \over q}, \\
		&	j_y^{(surf)} =  \chi_1 (E_yE_z^*+E_zE_y^*)  - \gamma P_{\rm circ} |E_0|^2 {q_x \over q}, \nonumber
\end{align}
where $P_{\rm circ}$ is the circular polarization degree given by ${P_{\rm circ} |E_0|^2 = {\rm i}(\bm E \times \bm E^*)\cdot \bm q/q}$. 
Here the constants $\chi_1$ and $\gamma$ describe the linear and circular PGE currents at oblique incidence which both are odd in the incidence angle $\Theta$.

The photon drag effect induced photocurrent is a sum of two terms
\begin{equation}
	\bm j^{(drag)} = \bm j^{(ax)} + \bm j^{(tr)}.
\end{equation}
The contribution $\bm j^{(ax)}$ is independent of the microscopic details of the studied system and has the same form as in any uniaxial system~\cite{GlazovGanichev_review}:
\begin{align}
j_x^{(ax)} =& \tilde{\cal T} P_{\rm circ} |E_0|^2 {q_y q_z \over q} + {\cal T}_1 q_x (|E_x|^2+|E_y|^2)  \\
&+ {\cal T}_2 \left[q_x (|E_x|^2-|E_y|^2) + q_y(E_xE_y^*+E_yE_x^*)\right]  \nonumber\\
& + {\cal T}_3 q_z (E_xE_z^*+E_zE_x^*) + {\cal T}_4 q_x |E_z|^2, \nonumber \\ 
j_y^{(ax)} = &-  \tilde{\cal T} P_{\rm circ} |E_0|^2 {q_x q_z \over q} + {\cal T}_1 q_y (|E_x|^2+|E_y|^2) \nonumber \\ 
&+ {\cal T} \left[q_x(E_xE_y^*+E_yE_x^*)-q_y (|E_x|^2-|E_y|^2)\right] \nonumber \\ 
&+ {\cal T} q_z (E_yE_z^*+E_zE_y^*)+ {\cal T} q_y |E_z|^2.   \nonumber
\end{align}
Here $\tilde{\cal T}$ describes the circular photon drag current, and ${\cal T}_{1\ldots 4}$ are constants of the linear photon drag effect.
Like $\bm j^{(surf)}$, the contribution $\bm j^{(ax)}$ is odd in the incidence angle $\Theta$ and at small small $\Theta$ depends linearly.

The second kind of the photon drag photocurrents, $\bm j^{(tr)}$, 
is specific for trigonal systems and exists for both C$_{\rm 3v}$ symmetry of the surface 
and D$_{\rm 3d}$ symmetry of the bulk (in both cases the reflection plane is $zx_0$):
\begin{align}
	j_{x_0}^{(tr)} =&{\cal T} [q_{x_0}(E_{x_0}E_z^*+E_{x_0}^*E_z)-q_{y_0}(E_{y_0}E_z^*+E_{y_0}^*E_z)] \nonumber \\
	&+ {\cal T}' P_{\rm circ} |E_0|^2 {2q_{x_0}q_{y_0}\over q},
\end{align}
\begin{align}
	j_{y_0}^{(tr)} =& -{\cal T} [q_{y_0}(E_{x_0}E_z^*+E_{x_0}^*E_z)+q_{x_0}(E_{y_0}E_z^*+E_{y_0}^*E_z)] \nonumber \\
	&+ {\cal T}'P_{\rm circ} |E_0|^2 {q_{x_0}^2-q_{y_0}^2\over q},
\end{align}
Note that, in contrast to  $\bm j^{(surf)}$ and $\bm j^{(ax)}$,  
 the trigonal contribution is an even function of the incidence 
angle: $j^{(tr)} \propto \Theta^2$ at small $\Theta$.

\section{Derivation of the photocurrent}

For derivation the expression for the photocurrent we solve the kinetic equation
\begin{equation}
    {\partial f_{\bm p} \over \partial t} + e \bm E \cdot {\partial f_{\bm p} \over \partial \bm p}
    = - \sum_{\bm p'} \left( W_{\bm p', \bm p} f_{\bm p} - W_{\bm p, \bm p'} f_{\bm p'} \right),
\end{equation}
up to the second order in the field $\bm E$ and in the first order in $W^{(a)}_{\bm p', \bm p}$. We search the electron distribution function in the form
$f_{\bm p}(t) = f_0(\varepsilon_p) + f_{\bm p}^{(1)}(t) +  f_{\bm p}^{(2)}$
with the oscillating in time term $f_{\bm p}^{(1)}(t)$
and the stationary term  $f_{\bm p}^{(2)} \propto |\bm E|^2$. The correction $f_{\bm p}^{(1)}$ is given by the standard expression
\begin{equation}
    f_{\bm p}^{(1)} = -e \tau_{1\omega} f_0' \bm E \cdot \bm v_{\bm p} + c.c.,
\end{equation}
where $f_0'=df_0/d\varepsilon_p$, and we introduce complex relaxation rates $\tau_{n\omega}^{-1} = \tau_n^{-1} - {\rm i}\omega$ with the elastic relaxation times ($n=1,2$) being
\begin{equation}
 {1\over\tau_n(\varepsilon_p)}=\sum_{\bm p'}W^{(s)}_{\bm p', \bm p}[1-\cos{n(\varphi_{\bm p'} - \varphi_{\bm p}})] .
\end{equation}

There are two contributions to $f_{\bm p}^{(2)}$. The first one is obtained by the following way: one first finds the quadratic in $\bm E$ correction $\widetilde{f_{\bm p}}$ satisfying the equation
\begin{equation}
- {\rm i}\omega \widetilde{f_{\bm p}}    + e \bm E \cdot {\partial f_{\bm p}^{(1)} \over \partial \bm p}
    = - \sum_{\bm p'} W^{(s)}_{\bm p', \bm p} \left( \widetilde{f_{\bm p}} - \widetilde{f_{\bm p'}} \right),
\end{equation}
and then find $f_{\bm p}^{(2)}$ from the equation
\begin{equation}
    \sum_{\bm p'} W^{(a)}_{\bm p', \bm p}  \widetilde{f_{\bm p'}} + {f_{\bm p}^{(2)} \over \tau_1} = 0.
\end{equation}
Here we used the property $\sum_{\bm p'} W^{(a)}_{\bm p', \bm p} =0$.

The second contribution is found by taking into account the scattering asymmetry and the second power of the field in the opposite way. In this approach, the oscillating $\bm E$-linear correction $\widetilde{f}_{\bm p}^{(1)}$ satisfies the equation
\begin{equation}
    \sum_{\bm p'} W^{(a)}_{\bm p', \bm p} f_{\bm p'}^{(1)} + {\widetilde{f}_{\bm p}^{(1)}\over \tau_{2\omega}} = 0,
\end{equation}
and then the contribution $\delta f_{\bm p}^{(2)}$ is found from
\begin{equation}
    e \bm E \cdot {\partial \widetilde{f}_{\bm p}^{(1)} \over \partial \bm p} = - {\delta f_{\bm p}^{(2)} \over \tau_1}.
\end{equation}

Calculating the electric current by
\begin{equation}
    \bm j = e\sum_{\bm p} \left(f_{\bm p}^{(2)}+\delta f_{\bm p}^{(2)} \right)v_0{\bm p \over p},
\end{equation}
we obtain the result as a sum of two contributions corresponding to  $f_{\bm p}^{(2)}$ and $\delta f_{\bm p}^{(2)}$, respectively:
\begin{align}
    j_x = &|E_x|^2 {N(ev_0)^3\over 1+(\omega\tau_{\rm tr})^2} \\
    &\times \left[
    {\tau_{\rm tr}\over \varepsilon_{\rm F}^3} {d (\Xi\tau_2\varepsilon_{\rm F}^2) \over d\varepsilon_{\rm F}}
    +{1-\omega^2\tau_{\rm tr}\tau_2\over 1+(\omega\tau_2)^2}
    \Xi\tau_2 {d(\tau_{\rm tr}/ \varepsilon_{\rm F})\over d\varepsilon_{\rm F}}
    \right], \nonumber
\end{align}
where $\tau_{\rm tr} = \tau_1(\varepsilon_{\rm F})$, $\tau_2$ also should be taken at $\varepsilon_p=\varepsilon_{\rm F}$, and the factor $\Xi$ is defined as
\begin{equation}
\label{Xi_suppl}
    \Xi = \tau_{\rm tr} \sum_{\bm p'}
    \left< 2\cos{\varphi_{\bm p}} \cos{2\varphi_{\bm p'}} \, W^{(a)}_{\bm p', \bm p} \right>_{\varphi_{\bm p}},
\end{equation}
where the brackets denote averaging over the directions of $\bm p$ at the Fermi circle.

\section{Estimation of ${\Xi}$}

The surface states in TI are formed in the $\Gamma$ point from atomic $z$-orbitals which are odd and even in respect to the space inversion~\cite{NatPhys09}. We denote the corresponding functions as $Z_-$ and $Z_+$, respectively, and the energy gap as $E_g$. The upper- and lower-lying bands are 
formed from odd ($X_-,Y_-$) and even ($X_+,Y_+$) $x$ and $y$ orbitals. 
For brevity, hereafter we assume that the Cartesian coordinates $x,y$ coincide with the crystallographic axes $x_0,y_0$.
With account for the spin-orbit splitting, the Bloch functions in these bands in the $\Gamma$ point are ${(X_- + {\rm i} Y_-) / \sqrt{2}}$ and ${(X_+ - {\rm i} Y_+) / \sqrt{2}}$. We denote the energy gaps between $Z_+$ and ${(X_- + {\rm i} Y_-) / \sqrt{2}}$ as $E_{g1}$, and between ${(X_+ - {\rm i} Y_+) / \sqrt{2}}$ and $Z_-$ as $E_{g1}'$.
In order to take into account mixing of the states at $p \neq 0$, we introduce the interband matrix elements of the momentum operator which are nonzero in the axial approximation:
\begin{equation}
    P = {{\rm i} \over m_0} \langle Z_-| p_x | X_+\rangle = {{\rm i} \over m_0} \langle Z_-| p_y | Y_+\rangle,
\end{equation}
\begin{equation}
    P' = {{\rm i} \over m_0} \langle Z_+| p_x | X_-\rangle = {{\rm i} \over m_0} \langle Z_+| p_y | Y_-\rangle,
\end{equation}
the matrix element which reflects the D$_{\rm 3d}$ symmetry of the bulk crystal:
\begin{align}
    R &= {{\rm i} \over m_0} \langle X_+| p_y | Y_-\rangle = {{\rm i} \over m_0} \langle Y_+| p_y | X_-\rangle \\
    & = {{\rm i} \over m_0} \langle Y_+| p_x | Y_-\rangle = - {{\rm i} \over m_0} \langle X_+| p_x | X_-\rangle, \nonumber
\end{align}
and the matrix element accounting for the absence of ${z \to -z}$ reflection in surface:
\begin{equation}
    V = \langle Z_+| H_{surf} |Z_-\rangle.
\end{equation}
Note that the values of $V$ have opposite signs at two opposite (ideal) surfaces of the crystal.

Up to the second order in $\bm p$ and the first order in $V$, the Bloch functions $Z_\pm(\bm p)$
at a finite momentum $\bm p$ have the following form:
\begin{align}
\label{Z+}
        Z_-(\bm p) &= Z_- + {1 \over \sqrt{2}(E_g + E_{g1})} \\
&   \times  \left[ {{\rm i}V P' (p_x - {\rm i}p_y) \over  E_g }
    - {R P (p_x + {\rm i}p_y)^2 \over  E_{g1}' }
    \right] { X_- + {\rm i}Y_- \over \sqrt{2}}, \nonumber
    \end{align}
\begin{align}
    \label{Z-}
    Z_+(\bm p) &= Z_+ + {1 \over \sqrt{2}(E_g + E_{g1}')} \\
&   \times  \left[ {{\rm i}V P (p_x + {\rm i}p_y) \over  E_g }
    + {R P' (p_x - {\rm i}p_y)^2 \over  E_{g1} }
    \right] { X_+ - {\rm i}Y_+ \over \sqrt{2}}. \nonumber
\end{align}

At $p \neq 0$, the surface states have the spinor wavefunctions
being coherent superpositions of $Z_-$ and $Z_+$ orbitals multiplied by the spin-up and spin-down spinors.
Taking into account the $\bm p$-dependence of the Bloch amplitudes Eqs.~\eqref{Z+},~\eqref{Z-} we obtain that
the matrix element of elastic scattering by a potential $U(\bm r)$
\[
U_{\bm p' \bm p} = U_0(\bm p' - \bm p)
     {\langle Z_-(\bm p')|Z_-(\bm p)\rangle + \langle Z_+(\bm p')|Z_+(\bm p)\rangle\over 2}
\]
contains the following term
\begin{align}
U_{\bm p' \bm p} &   \propto U_0(\bm p' - \bm p)\Biggl\{ 1 - {\rm i} {VRPP' \over 4E_g} \\
    \times  & \biggl[ {(p_x + {\rm i}p_y)^2(p'_x + {\rm i}p'_y) -
    (p'_x - {\rm i}p'_y)^2(p_x - {\rm i}p_y)  \over (E_g + E_{g1})^2  E_{g1}'}
     \nonumber\\
&       + { (p_x - {\rm i}p_y)^2(p'_x - {\rm i}p'_y) -
(p'_x + {\rm i}p'_y)^2(p_x + {\rm i}p_y)  \over (E_g + E_{g1}')^2 E_{g1}}
    \biggr] \Biggr\}. \nonumber
\end{align}
Here $U_0(\bm p' - \bm p)$ is the Fourier image of $U(\bm r)$ multiplied by the factor $\cos[({\varphi_{\bm p'} - \varphi_{\bm p}) /2 ]}$.
Note that a similar contribution can be obtained from admixture of $X_+ - {\rm i}Y_+$ to $Z_-$ and $X_- + {\rm i}Y_-$ to $Z_+$ wavefunction by $Vp^2$- and $p$-linear terms.

The presence of both real and imaginary terms in the scattering matrix elements leads to
the scattering probability in the form
\begin{equation}
    W_{\bm p', \bm p} = W^{(s)}_{\bm p', \bm p} + W^{(a)}_{\bm p', \bm p},
\end{equation}
where the symmetrical part is given by the standard expression
\begin{equation}
    W^{(s)}_{\bm p', \bm p} = {2\pi\over\hbar} {\cal N} \sum_{{\bm p}'}
    |U_0(\bm p-\bm p')|^2
    \delta(\varepsilon_p - \varepsilon_{p'})
\end{equation}
with ${\cal N}$ being a concentration of scatterers.
The antisymmetrical part of the scattering probability is obtained in the third order in the scattering potential~\cite{Belinicher-Strurman-UFN,Sturman-Fridkin-book}:
\begin{align}
     W^{(a)}_{{\bm p} {\bm p}'} = &- {2\pi\over\hbar}{\cal N}\delta(\varepsilon_p - \varepsilon_{p'}) \\
    &\times  2\pi \sum_{{\bm p}_1} {\rm Im}
    \left(U_{{\bm p} {\bm p}_1}U_{{\bm p}_1{\bm p}'}U_{{\bm p}' {\bm p}}\right)
 \delta(\varepsilon_p - \varepsilon_{p_1}). \nonumber
\end{align}
This yields the following estimate
\begin{align}
W^{(a)}_{{\bm p} {\bm p}'} &    \approx {(2\pi)^2\over\hbar}{\cal N}\delta(\varepsilon_p - \varepsilon_{p'}) {VRPP' p_{\rm F}^3 \over 4 E_g}
    \\
& \times \left[ {1\over (E_g + E_{g1})^2E_{g1}'} + {1\over (E_g + E_{g1}')^2E_{g1}}\right]
\nonumber\\
& \times    \sum_{{\bm p}_1}
    F(\varphi_{\bm p},\varphi_{\bm p'},\varphi_{\bm p_1}) \delta(\varepsilon_p - \varepsilon_{p_1}),
    \nonumber
\end{align}
where
\begin{align}
&   F(\varphi_{\bm p},\varphi_{\bm p'},\varphi_{\bm p_1}) =
    U_0({{\bm p}- {\bm p}_1})U_0({{\bm p}_1 - {\bm p}'}) U_0({{\bm p}' - {\bm p}}) \nonumber\\
&   \times [\cos{(2\varphi_{\bm p}+\varphi_{\bm p_1})} - \cos{(2\varphi_{\bm p_1}+\varphi_{\bm p})}
    +\cos{(2\varphi_{\bm p_1}+\varphi_{\bm p'})} \nonumber \\
&   - \cos{(2\varphi_{\bm p'}+\varphi_{\bm p_1})}   +\cos{(2\varphi_{\bm p'}+\varphi_{\bm p})} - \cos{(2\varphi_{\bm p}+\varphi_{\bm p'})} ]. \nonumber
\end{align}
For short-range scattering potential 
$U_0(\bm p'-\bm p)=u\cos{[(\varphi_{\bm p'}-\varphi_{\bm p})/2]}$ with $u$ being a constant we derive
 $\Xi=0$, so the PGE current is not generated. This is a specific feature of Dirac fermions: for massive particles with a parabolic energy dispersion, $U_0({{\bm p}' - {\bm p}})$ is independent of $\varphi_{\bm p'} $ and $ \varphi_{\bm p}$ for short-range scattering, and PGE is present.
For Coulomb impurity scattering, when ${U_0({{\bm p}' - {\bm p}}) \propto \cos{[(\varphi_{\bm p'} - \varphi_{\bm p}) /2 ]}/|\sin{[(\varphi_{\bm p'} - \varphi_{\bm p}) /2 ]}| }$,  $\Xi$ is finite, i.e. PGE is present for Dirac fermions as well as for massive electrons.

We can estimate the factor $\Xi$ defined in Eq.~\eqref{Xi_suppl} as
\begin{equation}
 \Xi \sim {\varepsilon_{\rm F} U_s \over (\hbar v_0)^2} \left( {\varepsilon_{\rm F} \over E_g}\right)^3 {V\over E_g} {RPP' \over v_0^3},
\end{equation}
where $U_s$ is the characteristic energy of the scattering potential multiplied by its characteristic area.
For ${\varepsilon_{\rm F} U_s / (\hbar v_0)^2 \sim 1 }$, $\varepsilon_{\rm F} / E_g \sim 10^{-2}$, $V/E_g \sim 10^{-1}$, $P \sim P' \sim v_0$, and $R/v_0 \sim 10^{-2}$ we obtain $\Xi \sim 10^{-5}$.
This value is consistent with the experimental results.